\documentclass[letter]{llncs}
\usepackage[margin=1.54in]{geometry}
\usepackage{amssymb,amsmath,authblk}
\usepackage{graphicx}
\usepackage{tikz}
\usepackage{color}

\usepackage{url}
\newcommand{\keywords}[1]{\par\addvspace\baselineskip
\noindent\keywordname\enspace\ignorespaces#1}

\newcommand{\bdf}{\begin{definition}}
\newcommand{\edf}{\end{definition}}
\newcommand{\beq}{\begin{equation}}
\newcommand{\eeq}{\end{equation}}
\newcommand{\bsp}{\begin{split}}
\newcommand{\esp}{\end{split}}

\newcommand{\EE}{\mathbb{E}}
\newcommand{\start}{\text{start}}
\newcommand{\finish}{\text{finish}}

\definecolor{lightgray}{rgb}{0.85,0.85,0.85}
\definecolor{mediumgray}{rgb}{0.75,0.75,0.75}
\definecolor{darkgray}{rgb}{0.45,0.45,0.45}

\newcommand\restr[2]{{
  \left.\kern-\nulldelimiterspace 
  #1 
  \vphantom{\big|} 
  \right|_{#2} 
  }}

\begin{document}

\mainmatter

\title{Thermodynamic Analysis of Classical and Quantum Search Algorithms}

\titlerunning{Thermodynamic Analysis of Search}


\author{Ray Perlner$^1$ \and Yi-Kai Liu$^{1,2}$}
\institute{$^1$National Institute of Standards and Technology (NIST)\\
Gaithersburg, MD, USA\\
$^2$Joint Center for Quantum Information and Computer Science (QuICS)\\
University of Maryland, College Park, MD, USA}

\maketitle

\begin{abstract}
We analyze the performance of classical and quantum search algorithms from a thermodynamic perspective, focusing on resources such as time, energy, and memory size. We consider two examples that are relevant to post-quantum cryptography: Grover's search algorithm, and the quantum algorithm for collision-finding. Using Bennett's ``Brownian'' model of low-power reversible computation, we show classical algorithms that have the \textit{same} asymptotic energy consumption as these quantum algorithms. Thus, the quantum advantage in query complexity does not imply a reduction in these thermodynamic resource costs. In addition, we present realistic estimates of the resource costs of quantum and classical search, for near-future computing technologies. We find that, if memory is cheap, classical exhaustive search can be surprisingly competitive with Grover's algorithm.

\end{abstract}


\section{Introduction}

\subsection{Motivation}

Quantum computers are believed to solve a number of important problems, in areas such as number theory, physical simulation and combinatorial search, asymptotically faster than classical computers \cite{NC00}. How large is this quantum speedup? There is now a rich body of literature that analyzes different quantum speedups, using idealized models of computation, such as the quantum circuit model, and quantum oracle models (i.e., quantum query complexity). In most of this work, the models of computation are intentionally made simple, in order to allow rigorous analyses; for instance, in quantum query complexity, one only accounts for the number of oracle queries made by the algorithm, while disregarding the actual time-complexity of the algorithm. 

In order to obtain more realistic analyses of quantum speedups, one must consider more realistic models of quantum computation, which take into account time-complexity (not just query complexity), as well as possible time-space tradeoffs due to the use of parallel processors, and restrictions on the connectivity of the different components of the computer. There have been a few results of this type \cite{bernstein2009cost,beals2013efficient,Fluhrer2017,Banegas2017}.  Some of these results suggest that practical quantum speedups may fall short of the most idealized theoretical predictions.

In this paper we give a new analysis of some fundamental quantum speedups, from the perspective of thermodynamics. A computer can be viewed as an engine that converts energy into computational work; and one may ask how much energy is consumed by running a particular algorithm. To answer this question, one must specify the model of computation. Following Bennett \cite{Bennett1982}, one can consider three different models:
\begin{enumerate}
\item ``Conventional'' computation (as in present-day electronic computers) uses operations that are irreversible and deterministic. Since the operations are irreversible, at temperature $T$, each operation must dissipate at least $k T (\ln 2)$ of energy, where $k$ is Boltzmann's constant. (This is the ``Landauer limit.'')
\item ``Ballistic'' computation (as in billiard ball models \cite{Fredkin1982}) uses operations that are reversible and deterministic. Since the operations are reversible, in principle, they can dissipate zero energy. The computer is assumed to be isolated from all sources of thermal noise, hence energy barriers are not needed to prevent errors.
\item ``Brownian'' computation (as in DNA computation \cite{Bennett1973} and adiabatic circuits \cite{arsalan2007asynchronous}) uses operations that are reversible and stochastic. Each operation dissipates a small amount of energy $\varepsilon$, which may be less than $k T$, so that the computation ``drifts'' forward, even in the presence of strong thermal noise. 
\end{enumerate}
By using reversible operations, models (2) and (3) are able to compute using an amount of energy per operation that is below the Landauer limit. However, it has been argued that model (2) is unrealistic, because it cannot be made fault-tolerant, i.e., it is sensitive to small errors when performing a long computation. In this paper, we use model (3), the Brownian model of computation, as our standard.

We consider quantum and classical algorithms for unstructured search and collision finding. Using the Brownian model of computation, we analyze the cost of these algorithms in terms of time, energy consumption, and memory size. The motivation for this study comes from post-quantum cryptography. For instance, an algorithm for unstructured search can be used to recover the secret key of a block cipher, given a sufficient number of plaintext-ciphertext pairs; while an algorithm for collision-finding can be used to compromise the security of a cryptographic hash function. In order to design block ciphers and hash functions that achieve sufficiently high levels of security, one must make detailed estimates of the resources required to carry out both quantum and classical cryptanalytic attacks.

\subsection{Our Results}

For the problem of collision finding, previous work suggested that quantum algorithms were unlikely to provide an asymptotic advantage in terms of circuit size (despite using fewer oracle queries) \cite{bernstein2009cost}. Our thermodynamic analysis leads to a similar conclusion. We compare in detail the classical collision finding algorithm of Van-Oorschot and Wiener \cite{vanOorschot1999}, and the quantum collision finding algorithm of  Brassard, H{\o}yer, and Tapp (BHT \cite{Brassard1998},) including parallelized generalizations of BHT. We find that the energy consumption required to search for collisions on a range of size $N$ using a memory of size $M < O(N)$ in time $t$ is $O\left(\frac{N}{Mt}\right)$, regardless of the choice of algorithm. 

While we focus on the collision finding problem, it should be noted that similar analysis may also be applied to the Claw Finding problem, which seeks to find collisions between two functions with domain sizes $N_1$ and $N_2$ . A quantum algorithm was proposed by Tani for this purpose \cite{Tani2007}. This may be compared to the algorithm given by Van-Oorschot and Wiener \cite{vanOorschot1999}. In this case, the energy consumption required to find a claw using a memory of size $M < O(N_1 + N_2)$ in time $t$ is $O\left(max\left(\frac{N_1N_2}{Mt},  \frac{N_1}{t}, \frac{N_2}{t}\right)\right)$ for the quantum algorithm, and $O\left(\frac{(N_1+N_2)^3}{M^2t}\right)$ for the classical algorithm.

For the problem of unstructured search, it was known previously that Grover's algorithm does achieve a quadratic speedup over classical exhaustive search, both in terms of circuit size, and in terms of oracle queries. Quite surprisingly, we do \textit{not} find a quantum advantage using our thermodynamic analysis. On the contrary, we find that a Brownian implementation of classical random search can achieve the \textit{same} asymptotic performance as Grover's algorithm (up to logarithmic factors), where we measure the performance in terms of running time, memory size and energy consumption. 

To show this, we use a variant of the Brownian model, where certain steps in the computation are \textit{unpowered}, in the sense that we set $\varepsilon = 0$, so that no energy is dissipated, and the computation is simply driven by random thermal noise, with equal probability of moving forwards or backwards. Energy consumption is instead dominated by memory initialization costs. This model may be of independent interest. Our analysis shows that, in order to find a preimage within a domain of size $N$ using a memory of size $M$ in time $t$, both Grover's algorithm and unpowered classical search require an energy consumption of $O(\frac{N}{t})$, regardless of the memory size. 

Finally, we turn to a more detailed comparison of Grover's algorithm and (powered and unpowered) classical search. Unlike the case for quantum versus classical collision search, here there are some plausible reasons why Grover's algorithm may be more efficient than classical search in practice. In particular, for unpowered preimage search, the independence of memory size and energy consumption relies on a heuristic assumption of scale invariance. If we remove this assumption, and instead assume that unpowered preimage search can only be efficiently implemented at a fixed temperature scale $T$, we find that unpowered preimage search is significantly more memory intensive than Grover's algorithm. Additionally, unpowered preimage search is less efficient than Grover's algorithm when oracle queries have a large memory complexity, although this is only a minor problem in the typical scenario where the memory complexity of oracle queries scales logarithmically with $N$. 

\subsection{Near-Future Computing Technologies}

We end by making some quantitative estimates, based on the hypothetical scaling of near-future computing technologies. This is necessarily somewhat speculative. Nonetheless, we argue that one can draw some rough conclusions regarding applications such as brute-force cryptanalysis of block ciphers, which takes on the order of $2^{80}$ or $2^{96}$ operations, and thus is comfortably in the asymptotic regime. By making rough calculations based on asymptotic scaling, one can draw some conclusions that do not depend too much on any particular type of qubit, or any particular scheme for quantum error correction.

Our results suggest that the cost of constructing computing hardware (e.g., memory and CPU's) is a major factor that determines the practical advantage of running Grover's algorithm (as compared to classical brute-force search). If memory and CPU's are cheap, then one can use classical search and still be competitive with Grover's algorithm, simply by building enormous data centers. This seems obvious, on a qualitative level. Our quantitative estimates show that, in fact, classical computing technology can do surprisingly well. 
In particular, it is possible to envision a possible future state of technology where unpowered classical preimage search could outperform both Grover search and powered classical search. Such a scenario could occur if memory costs could be brought very close to fundamental thermodynamic limits, but quantum computers could not be implemented without very low operating temperatures and expensive error correction.

This paper is organized as follows. In Sections \ref{sec-pbc} and \ref{Unpowered}, we describe the model of Brownian computation, and its unpowered variant. In Sections \ref{sec-collision} and \ref{sec-preimage}, we give simple analyses of quantum and classical algorithms for collision search and preimage search, focusing on energy costs and time-space tradeoffs. In Sections \ref{PandT} and \ref{sec-oracle-costs}, we investigate the cost of preimage search at constant temperature and power, and we estimate the cost of implementing the oracle that checks each solution. Finally, in Section \ref{sec-near-future}, we make some detailed estimates of quantum-versus-classical speedups for some hypothetical future computing technologies, and in Section \ref{sec-conclusion}, we conclude.


\section{Powered Brownian Computation}
\label{sec-pbc}

\begin{figure}
\includegraphics[width=5.5in]{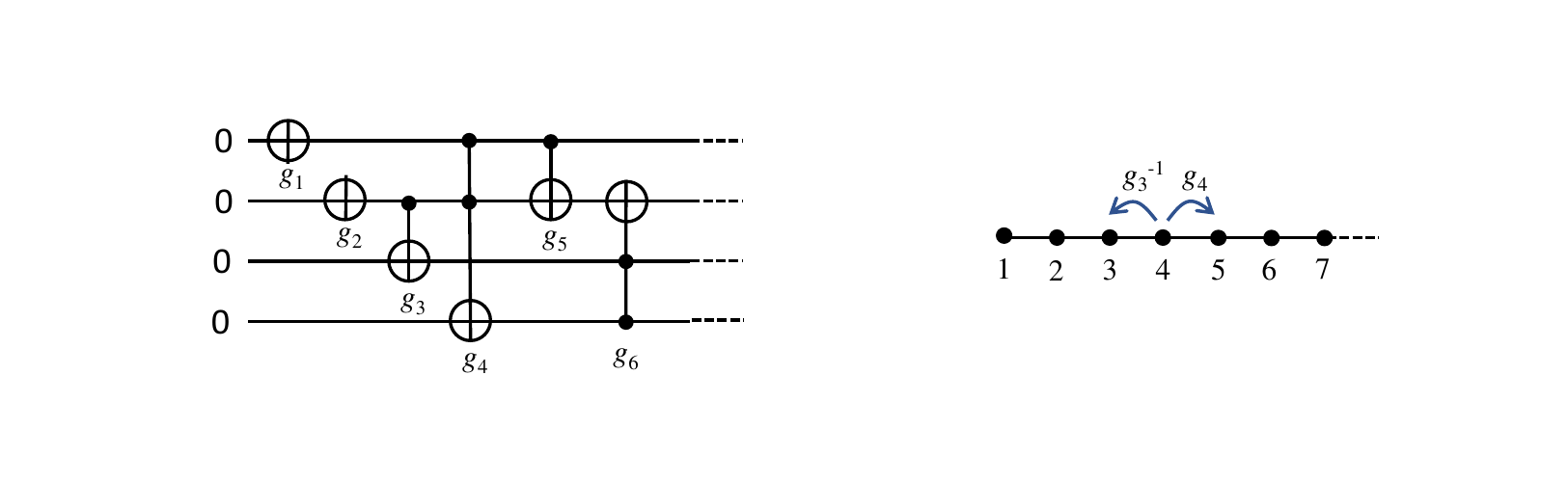}
\caption{A Brownian computation is described by a logically reversible circuit, whose gates are to be applied in some specified order. The dynamics of the Brownian computer are described by a random walk along a 1-dimensional chain.}
\label{fig-circuit}
\end{figure}

Brownian computers are assumed to operate near thermal equilibrium at a finite temperature, $T$. A program for a Brownian computer consists of a logically reversible circuit, whose gates (denoted $g_1,g_2,\ldots,g_m$) are to be applied in some specified sequential order. The time-evolution of the Brownian computer may be described as a random walk on a 1-dimensional chain, where the $i$'th vertex corresponds to the state of the computation after the first $i-1$ gates $g_1,g_2,\ldots,g_{i-1}$ have been applied. A forward step from vertex $i$ to vertex $i+1$ corresponds to applying the $i$'th gate $g_i$, and a backwards step from vertex $i+1$ to vertex $i$ corresponds to undoing the $i$'th gate $g_i$. (See Figure \ref{fig-circuit}.) In the absence of any driving force, forward and backward steps occur with equal probability. In order for a computation to proceed forward at a nonzero rate, a driving force, dissipating an energy of $\varepsilon$ per gate, is imposed. This causes forward steps to occur with $e^{\frac{\varepsilon}{kT}}$ times greater probability than backward steps, resulting in a net forward computation rate proportional to $\frac{\varepsilon}{kT}$, for $\varepsilon$ small compared to $kT$.

Note that the Brownian model of computation can be generalized to work with quantum circuits, provided that all operations are unitary, and all measurements are deferred to the end of the computation. Essentially, this amounts to running a quantum computer with a classical controller that behaves in a Brownian fashion.

\subsection{Energy Consumption and Running Time}

For the Brownian model, we can derive an asymptotic scaling for the relation between per-gate time and per-gate energy, assuming a fixed temperature $T$. Strictly speaking this analysis can only be extended to a range of possible temperatures under an assumption that physics is scale invariant within that range. However, we can give a somewhat heuristic argument specifying a lower bound, independent of temperature, on the per-gate energy $\varepsilon$ required to perform $G$ sequential operations in time $t$. Briefly, if we assume that Brownian motion within a reversible circuit can be modeled as a series of ballistic motions, each with typical energy scale $kT$ and each completing $O(1)$ gates, then we can apply the  Margolus-Levitin theorem \cite{Margolus1998} to bound the total rate at which forward and backward transitions occur in the circuit. This suggests that the rate at which gates are traversed due to Brownian motion is no more than $\frac{4kT}{h}$. Combining this with the expectation that approximately $G \cdot \frac{kT}{\varepsilon}$ total transitions are required to complete $G$ sequential operations, we obtain the bound: 

\[
t > G \cdot \frac{kT}{\varepsilon} \cdot \frac{h}{4kT} =  \frac{hG}{4\varepsilon}
\]

or equivalently:

\[
\varepsilon > \frac{hG}{4t}.
\]

As the above argument is somewhat heuristic, for the remainder of this paper we will ignore the small unitless factors in the above formula, and simply use $\varepsilon \sim \frac{\hbar G}{t}$ when we need to give a concrete estimate of the dissipation energy required to perform a serial computation at a desired rate. 

\subsection{Fault Tolerance}

Some additional costs are required in order for Brownian computation to be achieved fault-tolerantly. Energy barriers must be imposed to prevent transitions to physical states outside the reversible circuit, representing the prescribed computation path. In order to suppress the probability of such undesirable transitions so that a circuit of size $G$ can be completed with high probability, the size of these energy barriers must at least be on the order of $kT ln (\frac{kT}{\varepsilon}\cdot G).$ Additionally, dissipating a ``latching" energy of about $kT ln (\frac{kT}{\varepsilon})$ during the computation's final step is required to suppress backwards transitions once the computation has reached its halting state. These costs are described in detail in \cite{Bennett1982}.

The above costs may, however, be assumed to be negligible in a number of important cases: In particular, the latching energy will be negligible when $\frac{\varepsilon}{kT}$ is at least logarithmically more than $\frac{1}{G}$. Additionally, establishing energy barriers to non-computational paths is likely to be a negligible cost when a description of the circuit can be expressed in a physically compact form, for example, using looping constructs. More precisely, if we assume that the circuit can be compressed into a program with memory requirement $m_0$ (including both the memory required to store the program and the data it acts on), then the cost of imposing energy barriers should be on the order of $m_0\cdot kT ln (\frac{kT}{\varepsilon}\cdot G).$ This cost is negligible as long as $\frac{\varepsilon}{kT}$ is significantly larger than $e^{-\frac{G}{m_0}}$.

In fact, the initialization cost may be less than this, since it may be more proper to think of the initialization process as rearranging the energy barriers already present in the available raw materials for constructing our computer. The cost is then determined by the Landauer limit and the information content of the circuit, including appropriately large energy barriers. Since the size of these barriers does not need to be precislely specified, but merely bounded above $kT ln (\frac{kT}{\varepsilon}\cdot G)$, the information content of the circuit may grow sublogarithmically with $G$. All we can say with confidence is that the information content of the circuit is at least $m_0$, and therefore the initialization energy is at least on the order of $m_0\cdot kT$.

Finally, it is worth commenting on the feasibility of Brownian computation for quantum computers. Brownian computation was originally proposed as a way to improve the thermodynamic efficiency of classical computation. It should be noted that many of the techniques that have been proposed for fault tolerance in quantum compuation are thermodynamically irreversible, in particular, syndrome measurement and magic state preparation. These techniques cannot be used in a Brownian mode of computation. However, there are some proposed techniques, such as the use of Fibonacci anyons for universal quantum computation \cite{trebst2008short}, that may be able to achieve fault tolerance without requiring significant thermodynamic irreversibility (although even in such cases, the cost of fault tolerance is believed to be polylogarithmic in the size of the circuit\footnote{In addition to the need for logarithmic-size energy barriers, shared with the classical case, only a discrete subset of the continuous space of quantum gates can be implemented fault tolerantly in proposed systems. The remaining gates must be approximated, and the cost of doing this is believed to be logarithmic in the inverse of the approximation error. See e.g. \cite{Kliuchnikov2014}}.) We will therefore optimistically assume that quantum operations can be implemented in a Brownian fashion.


\section{Unpowered Brownian Computation}
\label{Unpowered}



For some of our results, we will use a variant of the Brownian model of computation, where the intermediate steps in the computation are \textit{unpowered}. More precisely, we dissipate energy when initializing the state of the computer, and when reading the final output; but for the intermediate steps in the computation, we set $\varepsilon = 0$, so that no energy is dissipated, and the computation has equal probability of moving forwards or backwards, driven by random thermal noise. (In other words, the computation is a random walk without any ``forward drift.'') We now describe this in more detail.

Formally, the computation is described by a random walk on a graph $G = (V,E)$, together with a marked vertex $v_{\start}$, a set of marked vertices $V_{\finish} \subset V$, and an energy threshold $\varepsilon_{\text{th}} > 0$. The computation proceeds as follows:
\begin{enumerate}
\item The computer initializes its memory. (This dissipates $\varepsilon_{\text{th}} s_{\text{max}}$ units of energy, where $s_{\text{max}}$ is the size of the computer's memory.) Then the random walk begins at $v_{\start}$. 
\item At every step, the walk moves from its current position $v$ to a neighboring vertex $w \in \Gamma(v)$ chosen uniformly at random. 
\item When the walk reaches a vertex $v$ that belongs to the set $V_{\finish}$, we say that the computation has returned a result, which consists of the vertex $v$. (To read out this result, the computer dissipates $\varepsilon_{\text{th}} \log_2 |V|$ units of energy.)
\end{enumerate}

We assume that the graph $G$, and the energy threshold $\varepsilon_{\text{th}}$, have a few specific properties. Then computations of this type can be implemented by the same physical mechanisms as in the usual Brownian model. Specifically, we make the following assumptions:
\begin{enumerate}
\item We assume that the graph $G$ has constant degree (say, at most 10), so that every step in the random walk can be implemented in constant time, by coupling the computer to a noisy environment. 
\item We assume that the energy threshold $\varepsilon_{\text{th}}$ is large enough so that auxiliary data stored in the computer's memory will remain stable for the duration of the computation. (In particular, this ensures that, once the random walk reaches a vertex in the set $V_{\finish}$, it will stay there for the remainder of the computation.)
\end{enumerate}

To illustrate this model of unpowered Brownian computation, we now consider some representative examples, and we analyze their energy consumption and running time.

\subsection{Energy Consumption}

First, consider an unpowered Brownian computation that uses a memory of size $s_{\text{max}}$, and runs in $\tau_{\text{max}}$ steps. We argue that the total energy consumption (call this $E$) grows almost linearly with $s_{\text{max}}$, but only \textit{logarithmically} with $\tau_{\text{max}}$. This is in contrast to powered Brownian computation, where $E$ grows linearly with the running time,\footnote{Note, however, that powered Brownian computation may have a shorter running time than unpowered Brownian computation; we will discuss this in the next section.} since energy is dissipated at every step. 

This can be seen as follows: Note that we have $E = O(s_{\text{max}} \varepsilon_{\text{th}})$. We need to choose $\varepsilon_{\text{th}}$ large enough so that errors will not occur in the computer's memory while it is running. We can estimate the probability of having an error (call this $p_{\text{err}}$) as follows: suppose that at temperature $T$, errors occur independently on each bit of the memory, at each step of the computation, with probability $\exp(-\varepsilon_{\text{th}} / k T)$. Then $p_{\text{err}}$ can be bounded by 
\begin{equation}
p_{\text{err}} \leq \tau_{\text{max}} s_{\text{max}} \exp(-\varepsilon_{\text{th}} / k T).
\end{equation}
For example, for any $c_0 > 0$, if we set the energy threshold $\varepsilon_{\text{th}}$ to be 
\begin{equation}
\varepsilon_{\text{th}} = k T( \ln(\tau_{\text{max}} s_{\text{max}}) + c_0 ), 
\end{equation}
then we have that $p_{\text{err}} \leq \exp(-c_0)$. Thus, in order to make $p_{\text{err}}$ small, it is sufficient to set $\varepsilon_{\text{th}}$ to grow logarithmically with $\tau_{\text{max}}$ and $s_{\text{max}}$. Furthermore, by increasing $\varepsilon_{\text{th}}$, we can force $p_{\text{err}}$ to drop exponentially.

\subsection{Running Time}

\begin{figure}
\includegraphics[width=5.5in]{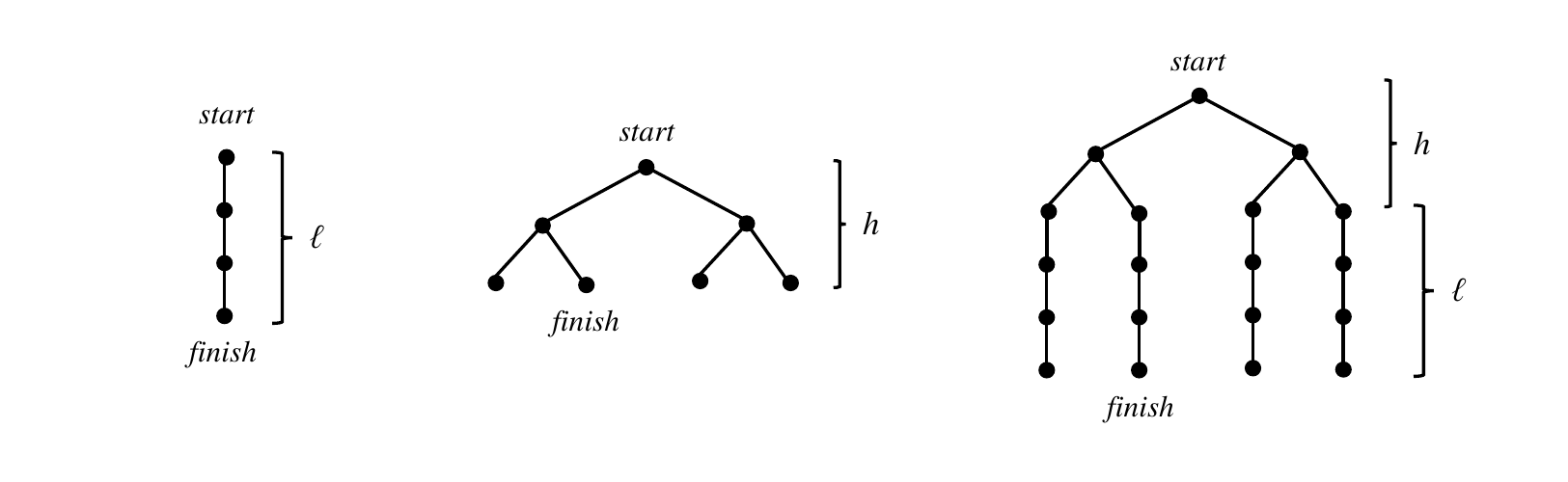}
\caption{Examples of unpowered Brownian computations: \textit{(left)} a sequential computation of length $\ell$, \textit{(middle)} a branching computation of depth $h$, \textit{(right)} a branching computation, followed by a sequential computation on each leaf.}
\label{fig-trees}
\end{figure}

We now consider three examples of unpowered Brownian computation (see Figure \ref{fig-trees}). We will compute the expected running times of these computations; that is, we let $\tau_{\finish}$ be the first time when the random walk reaches a vertex in $V_{\finish}$, and we compute the expected value $\EE(\tau_{\finish})$, averaging over the coin flips of the random walk. 

In general, we expect that unpowered Brownian computation will be slower than powered Brownian computation. However, this slow-down varies depending on the structure of the computation. In particular, we will show that sequential computations have a quadratic slow-down, whereas branching computations incur a slow-down that is only a logarithmic factor.

First, we consider a sequential computation of length $\ell$. Here, the graph $G$ is a chain of $\ell+1$ vertices. The expected running time of the computation is the expected time for a random walk to travel from one end of the chain to the other. A straightforward calculation \cite{Aldous} gives 
\begin{equation}
\EE(\tau_{\finish}) = \ell^2. 
\end{equation}
This is consistent with the intuition that a random walk on a 1-D chain will take $\sim r^2$ steps to move a distance $r$. Thus, for sequential computation, unpowered Brownian computation is quadratically slower than powered Brownian computation. 

Second, we consider a branching computation, which begins at the root of a binary tree of height $h$, and finishes when it hits a particular marked leaf of the tree. The expected running time can be upper-bounded as follows (Example 5.14 in \cite{Aldous}):
\begin{equation}
\EE(\tau_{\finish}) \leq 2(|V|-1)h < 4\cdot 2^h h. 
\end{equation}
For comparison, a deterministic search of the tree would take time $O(2^h)$. Hence, for branching computation, unpowered Brownian computation is only slightly slower than powered Brownian computation (they differ by a factor of $O(h)$, which is logarithmic in the total number of vertices). 

Finally, we will consider a computation that consists of $h$ branching steps, followed by $\ell$ sequential steps. Such a computation can be used to perform brute-force search: the computation first branches to select one of $2^h$ possible candidate solutions, then it does $\ell$ sequential operations to check whether that solution is correct. The expected running time can be upper-bounded as follows (Theorem 5.20 in \cite{Aldous}):
\begin{equation}
\EE(\tau_{\finish}) \leq 2(|V|-1)(h+\ell) < 2\cdot 2^h (\ell+2) (h+\ell).
\end{equation}
For comparison, a deterministic search of the tree would take time $O(2^h \ell)$. Hence, for brute-force search, this shows that unpowered Brownian computation is only slightly slower than powered Brownian computation, provided that $\ell \ll 2^h$ (i.e., one can quickly check whether a candidate solution is correct).

These observations suggest that branching computation (in the unpowered Brownian model) can be a useful tool for solving search problems. By using many parallel processors, one can make a natural time-space tradeoff. Moreover, in this situation, an unpowered Brownian algorithm can beat a powered Brownian algorithm, because its running time is not much worse (since most of the computation is branching rather than sequential), and its energy cost can be quadratically better (since the energy consumption scales linearly with space, but only logarithmically with time). As we will see in Section \ref{sec-preimage}, this can lead to classical search algorithms whose energy cost is competitive with Grover's algorithm.


\section{Collision Search}
\label{sec-collision}

The best known classical algorithm for finding collisions in a random function is the parallel collision search algorithm of Van Oorschot and Wiener \cite{vanOorschot1999}. If the range of the function is of size $N$, then, given $M$ parallel processes each with memory $O(1)$ the algorithm can find a collision in expected serial depth $O(\frac{\sqrt{N}}{M})$. The communication cost between threads is negligible compared to overall computational costs as long as $M$ is smaller than $\sqrt{N}$ by at least a logarithmic factor.

An improvement over classical collision search (and claw finding) has been claimed by Brassard, H{\o}yer, and Tapp (BHT) \cite{Brassard1998}. Their algorithm is a serial process consisting of $O(N^{\frac{1}{3}})$ operations and requires a memory of size $N^{\frac{1}{3}}$. This can be generalized to arbitrary memory size, $M<O(N^{\frac{1}{3}})$, giving a serial complexity of $O\left(\sqrt{\frac{N}{M}}\right)$. The BHT algorithm may be further generalized to a parallel algorithm involving $p$ parallel processors and a shared memory $M$, where $p<M<O\left((Np)^{\frac{1}{3}}\right)$.\footnote{Note this also implies that $M< O\left(\sqrt{N}\right)$. The constraint arises from the requirement that the serial complexity, $O\left(\frac{M}{p}\right)$, of filling a table of size $M$ with oracle values does not exceed the serial complexity $O\left(\sqrt{\frac{N}{Mp}}\right)$ of Grover search.} in this case, the serial complexity is $O\left(\sqrt{\frac{N}{Mp}}\right)$.

Bernstein \cite{bernstein2009cost} has observed that the BHT algorithm, even if parallelized, does not improve upon the Van Oorschot - Wiener algorithm, when measured in terms of memory and serial depth. Since the BHT algorithm also requires $O\left(\sqrt{\frac{N}{M}}\right)$ random access queries to a memory of size $M$, each requiring $O(M)$ gates, it also does not improve upon Van Oorschot Weiner algorithm when evaluated in terms of circuit size and depth (See Beals et al. \cite{beals2013efficient} for a more thorough analysis.) However, BHT does represent an improvement over all classical algorithms in terms of query complexity. Furthermore, the Quantum RAM model of Giovanetti et al. \cite{giovannetti2008quantum} gives a theoretical argument that despite their large gate complexity, quantum memory access operations can be performed at logarithmic energy cost. A question therefore remains whether there exists a physically realistic model of computation where BHT is actually cheaper than the classical algorithms for the same problem. However, if there is such a model, it is not the Brownian model of computation, as we proceed to show:

We first analyze the quantum algorithm, calculating the total energy required to perform a collision search, given a maximum time limit $t$ and a maximum memory size $M$. (Here, we assume, following the quantum RAM model, that the energy complexity of the BHT is dominated by oracle queries rather than memory access): The per operation energy $\varepsilon$ scales with the serial complexity divided by $t$, i.e.:

\begin{equation}\label{TBHT}
\varepsilon_{\text{quant}} = O\left(\frac{\sqrt{\frac{N}{Mp}}}{t}\right). 
\end{equation}
 
The total energy $E$ is then the product of the parallelism, the serial complexity, and the per operation energy, i.e.:

\begin{equation}\label{EBHT}
E_{\text{quant}} = O\left(p \cdot \sqrt{\frac{N}{Mp}} \cdot \frac{\sqrt{\frac{N}{Mp}}}{t} \right) = O\left(\frac{N}{Mt} \right).
\end{equation}

Now, we analyze the classical algorithm: The per operation energy again scales with the serial complexity, i.e.:

\begin{equation}\label{TVOW}
\varepsilon_{\text{cl}} = O\left(\frac{\sqrt{N}}{Mt}\right). 
\end{equation}

The total energy $E$ is again the product of the parallelism (In this case $p = O(M)$), the serial complexity, and the per operation energy, i.e.:

\begin{equation}\label{EVOW}
E_{\text{cl}} = O\left( M \cdot \frac{\sqrt{N}}{M} \cdot \frac{\sqrt{N}}{Mt} \right) = O\left(\frac{N}{Mt} \right).
\end{equation}

Thus, even under optimistic assumptions within the Brownian model of computation, we find that \textit{quantum computers provide no advantage in terms of energy, memory, or time, for solving the collision search problem}.  


\section{Preimage Search}
\label{sec-preimage}

Grover's algorithm finds preimages in a function with domain size $N$ in serial complexity $O(\sqrt N)$. Grover's algorithm can be generalized to take advantage of $M$ parallel processes each with memory $O(1)$, in which case the serial complexity is reduced to $O\left(\sqrt{\frac{N}{M}}\right)$. This serial complexity was shown to be optimal by Zalka \cite{Zalka1999}. If we implement Grover's algorithm in a Brownian fashion, we find that

\begin{equation}\label{TGrover}
\varepsilon_{\text{quant}} = O\left(\frac{\sqrt{\frac{N}{M}}}{t} \right),
\end{equation}

and,

\begin{equation}\label{EGrover}
E_{\text{quant}} = O\left(M \cdot \sqrt{\frac{N}{M}} \cdot \frac{\sqrt{\frac{N}{M}}}{t} \right) = O\left(\frac{N}{t}\right).
\end{equation}

A na\"ive Brownian implementation for classical search would divide the key space among $M$ parallel processes, each of which would deterministically step through$\frac{N}{M}$ keys searching for the correct one. Such a deterministic classical algorithm would require,

\begin{equation} \label{Tpowered}
\varepsilon_{\text{det}} = O\left(\frac{N}{Mt}\right), 
\end{equation}

and,

\begin{equation} \label{Epowered}
E_{\text{det}} = O\left( M \cdot \frac{N}{M} \cdot \frac{N}{Mt} \right) = O\left(\frac{N^2}{Mt} \right).
\end{equation}

This already allows us to compete with Grover's algorithm if we allow ourselves a memory of size $O(N).$
However, we can exploit the structure, or rather the lack of structure, of the search problem to improve upon this figure. In particular, rather than deterministically stepping through the keys, dissipating a driving energy each time, we can simply allow Brownian motion to drive the system on a random walk through the keyspace. (That is to say, we can use the unpowered Brownian computation model described in detail in section~\ref{Unpowered}.) We will still require a latching energy to end the computation, once the correct key has been found, and an initialization energy to create the necessary energy barriers to prevent unwanted transitions from occuring.

If the search is implemented by $M$ parallel processes, each of size $O(1)$, then each process must reach $\frac{N}{M}$ keys.  This requires the processes to operate at a temperature:

\begin{equation}\label{Tpreimage}
kT = O\left(\frac{N}{Mt}\right).
\end{equation}

The initialization energy should be of order $MkT$ i.e.:

\[
E_{\text{init}} = O\left(M \cdot \frac{N}{Mt}\right) =  O\left(\frac{N}{t}\right).
\]

This is identical to the energy required by a Brownian implementation of Grover's algorithm. All that remains is to show that the latching energy is negligible. Indeed, we find that the energy required to suppress backwards transitions from the final state for a time of order t is $O\left(kT ln(t kT)\right) = O\left(\frac{N}{Mt}ln\left(\frac{N}{M}\right)\right)$. This is negligible as long as $M$ is at least logarithmic in $N$.

Thus, as with collision search, \textit{the quantum and classical algorithms for preimage search appear to offer the same tradeoffs between time, energy and space}:
\begin{equation} \label{Epreimage}
E_{\text{cl}} = O \left(\frac{N}{t}\right) \text{ and } E_{\text{quant}} = O \left(\frac{N}{t}\right).
\end{equation}


\section{Preimage Search at Constant Power and Temperature}
\label{PandT}

In contrast to the collision search case, matching the time/ memory/ energy tradeoffs of Grover's algorithm with a classical search requires a somewhat unrealistic assumption. We assume that  if a computational process can be accomplished at a temperature $T$ in a time $t$, then an isomorphic computation can also be accomplished at a temperature $\alpha T$ in a time $\frac {T}{\alpha}$. This would be true if physics were scale invariant, but the physics of the real world is almost certainly not scale invariant. A more realistic model would therefore restrict the range of temperatures where a given computation is considered feasible. We will therefore repeat the analysis of the previous section assuming a fixed temperature $T$. For added realism, in addition to memory $M$, and time $t$, we will express the resources required for search in terms of power, $P = \frac{E}{t}$, rather than energy, since a fixed power budget is a more common limitation than a fixed energy budget.

From Equation~(\ref{Epreimage}) we find:
\[
N=O\left(Pt^2\right).
\]

Plugging this into Equation~(\ref{Tpreimage}) gives us:
\[
M=O\left(\frac{Pt}{T}\right).
\]

We can now calculate time and memory requirements in terms of $T$, $P$, and $N$:

\begin{align}\label{tpreimage}
t_{\text{cl}} = O\left(\sqrt{\frac{N}{P}}\right);\\
\label{Mpreimage}
M_{\text{cl}} = O\left(\frac{\sqrt{NP}}{T}\right).
\end{align}

A similar analysis may be done in the quantum case. Here we use Equation~(\ref{TGrover}) as a lower bound for $T$. If the per gate energy $\epsilon$ exceeds $kT$, we enter the thermodynamic regime of irreversible computing, as opposed to Brownian computing, at which point the time per gate not only fails to further decrease with increasing $\varepsilon$, but must in fact increase to prevent the waste heat from heating the computing system to a temperature higher than $T$. Combining this bound with Equation~(\ref{EGrover}) then yields the following time and memory requirements for Grover search at fixed power and temperature:

\begin{align}\label{tGrover}
t_{\text{quant}} = O\left(\sqrt{\frac{N}{P}}\right);\\
\label{MGrover}
M_{\text{quant}} = O\left(\frac{P}{T^2}\right).
\end{align}

Thus, fixing power and temperature, \textit{we find that our classical search strategy recovers the square-root time scaling of Grover's algorithm}. However, unlike Grover's algorithm, whose space requirement is determined only by the power budget and maximum operating temperature, \textit{the classical algorithm also requires memory} that scales, like the time, with the square root of the size of the search space.


\section{The Cost of Oracle Queries}
\label{sec-oracle-costs}

The asymptotic complexities given in previous sections ignore the computational complexity of individual oracle queries. Most of the results of previous sections remain substantively similar if these factors are included. We will model each oracle query as a circuit with depth $d_0$, width $m_0$, and total gates $g_0$.

In the case of powered Brownian computation, the effect of these factors is fairly straightforward. The memory imposed limit on parallelism (and number of table entries in the case of BHT) is now $p_{\text{max}} = O\left(\frac{M}{m_0}\right)$. Likewise, if $t_0$ is the time per query required to complete the computation in time $t$, we will now require an energy per gate of $\varepsilon = O\left(\frac{d_0}{t_0}\right)$. We must also ensure that all the bits or qubits in the circuit advance through it roughly synchronously. This can be done, for example, by associating a clock state of size $O\left(\log(d_0)\right)$ to each bit or qubit in the oracle circuit, and imposing a restoring potential proportional to the squared difference of the clock states of neighboring qubits. This will tend to couple the clock states of nearby qubits, but will not dissipate any net energy. As with other energy barriers ensuring correct computation, this potential need only extend logarithmically far from the equilibrium point, relative to the total size of the computation. We will generally ignore the logarithmic memory cost of the clock state and the logarithmic computational costs associated with creating interactions between the clock state, but in more detailed models, they may be subsumed into $m_0$ and $g_0$ respectively. Finally, we must take into account the number of gates required to perform an oracle query, $g_0$. 

\subsection{Collision search}

Making these substitutions into equations~(\ref{EBHT}) and (\ref{EVOW}) gives the following energy costs for quantum and classical collision search:

\begin{equation}\label{EBHTfull}
E_{\text{quant}} = O\left(p \cdot  g_0\sqrt{\frac{m_0N}{Mp}} \cdot d_0\frac{\sqrt{\frac{m_0N}{Mp}}}{t} \right) = O\left(\frac{g_0m_0d_0N}{Mt} \right);
\end{equation}

\begin{equation}\label{EVOWfull}
E_{\text{cl}} = O\left(\frac{M}{m_0} \cdot g_0\frac{m_0\sqrt{N}}{M} \cdot \frac{m_0d_0\sqrt{N}}{Mt} \right) = O\left(\frac{g_0m_0d_0N}{Mt} \right).
\end{equation}

Again, we find the classical and quantum complexities to be identical, up to constant factors. In both cases, the useful memory size is bounded above by $O\left(m_0\sqrt{N}\right)$.

\subsection{Preimage search}

Similarly, we may make the same substitutions in equations~(\ref{TGrover}) and (\ref{EGrover}) to include these factors in the per-gate and total energy cost of Grover search:

\begin{equation}\label{TGroverfull}
\varepsilon_{\text{quant}} = O\left(\frac{d_0\sqrt{\frac{m_0N}{M}}}{t} \right);
\end{equation}

\begin{equation}\label{EGroverfull}
E_{\text{quant}} = O\left(\frac{M}{m_0} \cdot g_0\sqrt{\frac{m_0N}{M}} \cdot \frac{d_0\sqrt{\frac{m_0N}{M}}}{t} \right) = O\left(\frac{g_0d_0N}{t}\right).
\end{equation}

In the case of unpowered Brownian computation, we must calculate the temperature $T$ required for random Brownian motion to power the traversal of an oracle circuit of depth $d_0$ and containing $g_0$ gates in time $t_0$. To do this, we create a random variable, $x$ indicating the total number of gates that have been completed at a time $t$. We expect that $x$ will obey the usual formula for Brownian motion, $\langle x^2 \rangle = Dt$, for some $D$, which will depend on $T$, $g_0,$ and $d_0$. We will then require $Dt_0 = O\left(g_0^2\right)$. It remains to determine the scaling of $D$. Note that at any given time, on average $O\left(\frac{g_0}{d_0}\right)$ gates will be exposed to activation by thermal noise. (The remaining gates will be disallowed by the clock states associated with their input/output bits.) Each of these gates is expected to contribute $O(Tdt)$ to $d\langle x^2\rangle$. The coupling potential between neighboring clock states will also drive the activation of individual gates, but it should have no net effect on $x$, since every gate driven forward by the coupling potential will be counterbalanced by another gate driven backwards. Thus we find that $D = O\left(\frac{Tg_0}{d_0}\right)$ and therefore $T = O\left(\frac{g_0d_0}{t_0}\right)$.

We may now apply this analysis to equations~(\ref{Tpreimage}) and (\ref{Epreimage}). Since, in order to complete a preimage search of size $N$ in time, $t$ with memory $M$, we need $t_0 = \frac{m_0N}{Mt}$, we find that:

\begin{equation}\label{Tpreimagefull}
T_{\text{cl}} = O\left(\frac{g_0m_0d_0N}{Mt}\right),
\end{equation}

and,

\begin{equation}\label{Epreimagefull}
E_{\text{cl}} = O\left(M \cdot \frac{g_0m_0d_0N}{Mt}\right) =  O\left(\frac{g_0m_0d_0N}{t}\right) = O\left(m_0 E_{\text{quant}}\right).
\end{equation}

Note that, when we include cost factors associated with the size and computational complexity of oracle queries, \textit{the mostly unpowered randomized preimage search is more energy intensive than Grover's algorithm by a factor of $O(m_0)$}. Nonetheless, this factor is generally expected to be logarithmic in $N$, and may easily be overwhelmed by the various costs associated with implementing fault tolerant quantum computation. 

Finally, we may also consider the fixed power and temperature scenario discussed in Section \ref{PandT}. In this case, equations~(\ref{tpreimage}), (\ref{Mpreimage}), (\ref{tGrover}) and (\ref{MGrover}) become:

\begin{align}
t_{\text{cl}} = O\left(\sqrt{\frac{g_0m_0d_0N}{P}}\right),\\
M_{\text{cl}} = O\left(\frac{\sqrt{g_0m_0d_0NP}}{T}\right),
\end{align}

and,

\begin{align}
t_{\text{quant}} = O\left(\sqrt{\frac{g_0d_0N}{P}}\right),\\
M_{\text{quant}} = O\left(\frac{m_0d_0P}{g_0T^2}\right).
\end{align}

For completeness, we will also consider the case of powered preimage search. Adapting equations~(\ref{Tpowered}) and (\ref{Epowered}) gives us:

\begin{equation}\label{Tpoweredfull}
\varepsilon_{\text{det}} = O\left( \frac{m_0d_0N}{Mt} \right),
\end{equation}

and,

\begin{equation}\label{Epoweredfull}
E_{\text{det}} = O\left(\frac{M}{m_0} \cdot g_0\frac{m_0N}{M} \cdot \frac{m_0d_0N}{Mt} \right) = O\left(\frac{g_0m_0d_0N^2}{Mt} \right).
\end{equation}

Note that, by comparing equations~(\ref{Epoweredfull}) and (\ref{EVOWfull}), we can see that the cost of powered preimage search with a domain of size $N$ is identical to the cost of collision search on a range of size $N^2$. 


\section{Near-Future Computing Technologies and the Grover Speedup}
\label{sec-near-future}

We are now in a position to estimate the practical relevance of Grover's algorithm and its classical counterpart, unpowered Brownian preimage search. The particular questions we intend to answer are the following: Is it reasonable, given Grover's algorithm (and its classical counterpart), to treat finding a preimage within a domain of size $N$ as an easier problem than finding a collision within a range of size $N^2$? How much easier? How do the answers to these questions depend upon the present and future state of technology -- in particular how do they depend upon the various ways that present and future technology may fall significantly short of thermodynamically ideal behavior?

The technology-dependent costs we will consider are: 

\begin{enumerate}
\item The cost of memory. For example, if we assume that power costs 10 cents per kWh and memory costs \$100 per TB, then the cost of a bit of memory is on the order of $memcost = 10^{15}kT$, where the temperature, $T$ is taken to be on the order of 300K. 
\item The increase in physical quantum circuit depth and gate count due to quantum error correction. Based on \cite{jones2012layered} we roughly estimate that near-future quantum error correction may increase memory requirements by a factor of $\frac{m_{\text{quant}}}{m_0} = 10^5$, effective circuit depth by a factor of $\frac{d_{\text{quant}}}{d_0} = 10^3$, and effective gate count by a factor of $\frac{g_{\text{quant}}}{g_0} = 10^8.$ 
\item The need for various quantum computing technologies to operate at extremely low temperatures. 
In addition to placing a lower limit on gate times, such low temperatures impose an energy cost due to the fact that any energy dissipated as heat at the lower temperature must eventually be removed and expelled to a heat bath, which typically must be at a much higher temperature, e.g. 300K. Moving heat from a sytem at a low temperature $T_{\text{quant}}$ to a system at a higher temperature $T$ increases energy consumption by a factor of $\frac{T}{T_{\text{quant}}}$.
\end{enumerate}

In the remainder of this section, we will study the relative cost of \textit{Grover's algorithm}, \textit{unpowered classical search}, and \textit{powered classical search}, when the above-mentioned technology-dependent cost factors take on our estimated current and near future values, and as they approach unity. For concreteness, we will also need to set values for the memory, $m_0$, circuit depth, $d_0$, and total gate count $g_0$ involved in oracle queries. Based on \cite{Grassl2016} we will estimate typical ranges for these values as $m_0 = 10^3,$ $d_0 = 10^5$ and $g_0 = 3 \times 10^6$.

 For each of the three algorithms considered, we will express the efficiency of the algorithm based on the maximum value of $N$ such that the search problem can be solved given an energy budget $E$ and a time budget $t$. We will denote this by $N_{\text{quant}}$, $N_{\text{det}}$ and $N_{\text{cl}}$, for Grover's algorithm, powered classical search, and unpowered classical search, respectively. 

We will take the time budget to be 1 year. As we will calculate relative efficiencies (i.e. $\frac{N_{\text{quant}}}{N_{\text{det}}}$ and $\frac{N_{\text{cl}}}{N_{\text{det}}}$), and in all cases $N$ will scale with $E$, we will not need to set a concrete value for $E$. Rather than providing an explicit memory budget, we will assume that the memory budget, $M$ is set so that $memcost \cdot M \leq E$.\footnote{Strictly speaking a larger memory budget is possible, since memory costs can be amortized accross multiple computations using the same hardware in series, but we judge that 1 year is a long enough time window to make this cost savings of only minor consideration.} By taking the log base 2 of these ratios, we can calculate Grover speedups in terms of the ''bits of security" metric typically used to evaluate cryptographic hardness. 

\subsection{Powered classical search}

We will first evaluate powered classical search (either for a collision within a range of size $N_{\text{det}}^2$ or for a preimage within a domain of size $N_{\text{det}}$.) We first solve for $M$ by setting $memcost \cdot M \leq E$, where the relation between $E$ and $N_{\text{det}}$ is given by Equation~(\ref{Epoweredfull}). We find that

\begin{equation}\label{MdetConcrete}
M = O\left(N_{\text{det}}\cdot \sqrt{\frac{\hbar g_0m_0d_0}{kTt}}\cdot\left(\frac{memcost}{kT}\right)^{-\frac{1}{2}}\right).
\end{equation}

Plugging $M$ back into Equation~(\ref{Epoweredfull}), and solving for $N_{\text{det}}$, we get:

\begin{equation}\label{Ndet}
N_{\text{det}} = O\left(E \cdot \sqrt{\frac{t}{\hbar g_0m_0d_0kT}}\cdot\left(\frac{memcost}{kT}\right)^{-\frac{1}{2}}\right)
\end{equation}

Note that the above computations assume a Brownian model of computation. To check that this is reasonable we must verify,  that $\varepsilon \leq kT$ for the range of values we're interested in. We will check this using equation~(\ref{Tpoweredfull}) with $M$ given by equation~(\ref{MdetConcrete}). Here we find:

\[\frac{\varepsilon}{kT} = O\left(\frac{\hbar m_0d_0N}{kTMt}\right) = O\left(\sqrt{\frac{\hbar m_0d_0}{g_0kTt}}\cdot \left(\frac{memcost}{kT}\right)^{\frac{1}{2}}\right). \]

If we use $\frac{memcost}{kT} = 10^{15}$ along with our other estimated values: $m_0 = 10^3$; $d_0= 10^5$; $g_0 = 3 \times 10^6$; $T= 300K$; $t = 1$ year, we get 
\[ \frac{\varepsilon}{kT} \approx 5\times 10^{-2}. \]
Letting  $\frac{memcost}{kT}$ approach unity yields 
\[ \frac{\varepsilon}{kT} \approx 2\times 10^{-9}. \]
As both values are less than 1, this indicates that the Brownian model of computation should yield a plausible, if slightly optimistic, approximation of $N_{det}$ accross the range of values of $memcost$ of interest to us. While these estimates of  
$\frac{\varepsilon}{kT}$ indicate that there may be advantages to using reversible computing even with current memory costs, it is not surprising that these advantages have not yet been realized, even for applications like bitcoin mining, which seems like a good fit but nonetheless continues to use standard irreversible computing technology at the time of writing. $5\times 10^{-2}$ does not differ from 1 by many orders of magnitude, and it could easily be overwhelmed by engineering costs not included in our model (fixed overhead associated with using reversible logic, extra gates required to synchronize different parts of a non-serial reversible circuit, different gate technology etc.) Nonetheless, it seems likely that if memory costs continue to fall, a broadly Brownian approach to computing will become cost effective. $2\times 10^{-9}$ differs from 1 by a significantly larger amount, and is less likely to be overcome by these sorts of overheads.

\subsection{Grover's algorithm}

We will now evaluate Grover's algorithm. Here all we need to do is modify Equation~(\ref{EGrover}) to account for technology-dependent cost factors and solve for $N_{\text{quant}}$.

\begin{equation}\label{NGrover}
N_{\text{quant}} = O\left( E \cdot \frac{t}{\hbar g_0d_0} \cdot \frac{T_{\text{quant}}}{T} \cdot \frac{g_0}{g_{\text{quant}}} \cdot \frac{d_0}{d_{\text{quant}}} \right).
\end{equation}

We compare Grover's algorithm with powered classical search, and we compute a speedup factor:

\begin{equation}\label{GroverSpeedup}
\frac{N_{\text{quant}}}{N_{det}} = O\left( \sqrt{\frac{m_0kTt}{\hbar g_0d_0}} \cdot \left(\frac{memcost}{kT}\right)^{\frac{1}{2}} \cdot \frac{T_{\text{quant}}}{T} \cdot \frac{g_0}{g_{\text{quant}}} \cdot \frac{d_0}{d_{\text{quant}}}\right).
\end{equation}

Using our estimated near future values for the technology dependent cost factors: $\frac{memcost}{kT} = 10^{15}$;  $\frac{m_{\text{quant}}}{m_0} = 10^5$; $\frac{d_{\text{quant}}}{d_0} = 10^3$; and $\frac{g_{\text{quant}}}{g_0} = 10^8$ along with our other estimated values:  $m_0 = 10^3$; $d_0= 10^5$; $g_0 = 3 \times 10^6$; $T= 300K$; $t = 1$ year, we get 
\[ \log_2\left(\frac{N_{\text{quant}}}{N_{\text{det}}}\right)\approx 4, \] 
indicating that Grover's algorithm will provide little, if any, advantage over classical search in the near future. Setting all technology-dependent cost factors to unity, yields a somewhat larger, but still modest, advantage of 
\[ \log_2\left(\frac{N_{\text{quant}}}{N_{\text{det}}}\right)\approx 21. \] 
In order to envision a larger advantage for Grover's algorithm, we must instead envision a scenario where classical memory remains as expensive as it is today, but all technology-dependent cost factors associated with quantum computers are eliminated. In this case, we get 
\[ \log_2\left(\frac{N_{\text{quant}}}{N_{\text{det}}}\right)\approx 46. \]

In the above analysis, we have ignored memory initialization costs for Grover's algorithm. To demonstrate that memory initialization costs do not necessarily overwhelm computation costs, we evaluate the memory requirements for Grover's algorithm at $T_{quant} = 10 mK$, assuming that physical qubits can be manufactured as cheaply as classical bits today (i.e., $memcost = 10^{15}kT$, for $T = 300K$). For fault-tolerance related cost factors, we make no special assumptions other than that error correction does not appreciably change circuit density (i.e., $\frac{g_{\text{quant}}}{m_{\text{quant}}d_{\text{quant}}} \approx \frac{g_0}{m_0d_0}$). Suitably modifying equation (\ref{TGroverfull}) gives us:

\[
\frac{memcost \cdot M_{\text{quant}}}{E} = O\left(\frac{\hbar m_0 d_0}{g_0kTt} \cdot \frac{m_{\text{quant}}}{m_0}\cdot \frac{d_{\text{quant}}}{d_0} \cdot \frac{g_0}{g_{\text{quant}}} \cdot \frac{T}{T_{\text{quant}}} \cdot \frac{memcost}{kT}\right) \approx 1.
\]

We also note that, while we have considered the possibility that quantum fault tolerance might be implemented in a fashion that avoids irreversible operations like measurement, if it cannot, this does not affect the above analysis. The energy cost of Grover's algorithm, aside from initialization, does not depend on memory within the Brownian computation regime. Thus, in order to minimize the memory requirement it is optimal to set $\varepsilon \approx kT$, at which point Brownian computation exhibits essentially the same energy costs as irreversible computation.

\subsection{Unpowered classical search}

Finally, we evaluate unpowered classical preimage search. Modifying Equation~(\ref{Epreimage}) gives us:

\begin{equation}\label{NPreimage}
N_{\text{cl}} = O\left( E \cdot \frac{t}{\hbar g_0m_0d_0} \cdot \left(\frac{memcost}{kT}\right)^{-1} \right).
\end{equation}

Again, we can compare this with powered classical search, by computing a speedup factor:
\begin{equation}\label{PreimageSpeedup}
\frac{N_{\text{cl}}}{N_{\text{det}}} = O\left( \sqrt{\frac{kTt}{\hbar m_0g_0d_0}} \cdot \left(\frac{memcost}{kT}\right)^{-\frac{1}{2}} \right)
\end{equation}

As before, we may use $\frac{memcost}{kT} = 10^{15}$ along with our other estimated values: $m_0 = 10^3$; $d_0= 10^5$; $g_0 = 3 \times 10^6$; $T= 300K$; $t = 1$ year. With these values, we get 
\[ \log_2\left(\frac{N_{\text{cl}}}{N_{\text{det}}}\right)\approx -14, \]
that is to say we find that unpowered classical search has a modest disadvantage over powered classical search assuming near future memory costs. However, this can be turned into a modest advantage of 
\[ \log_2\left(\frac{N_{\text{cl}}}{N_{\text{det}}}\right)\approx 11, \] 
if $\frac{memcost}{kT}$ goes to unity. A somewhat larger advantage may also be possible if we consider computations that last significantly longer than a year or scenarios where the ratio $\frac{memcost}{kT}$ reaches its optimum value at a temperature higher than $300K$.

In considering extremely optimistic scenarios for unpowered classical preimage search, it is worth noting that memory costs may be determined by the scarcity of matter rather than energy. We may estimate the total energy budget of the Earth, based on the total solar irradiance received by the Earth's atmosphere, which has been estimated at 174 PW \cite{174PW}. This translates to approximately $2 \times 10^{45} kT$ per year at 300K. In contrast, we may give an estimated matter budget based on the total number of atoms in the earth, which has been estimated around $10^{50}$ \cite{FermiAtoms}. As $10^{50}$ is a few orders of magnitude larger than $2 \times 10^{45}$, it remains plausible that energy could be the limiting factor in determining memory requirements, even given an extremely energy efficient manufacturing process, but the numbers are quite similar (especially if we are limited to only use atoms in the Earth's crust, for example.)


\section{Conclusion}
\label{sec-conclusion}

The development of quantum computing has created a great deal of excitement, particularly due to the discovery of quantum algorithms, such as Shor's algorithm, that perform exponentially better than the best known classical algorithm. Nonetheless, a large body of research concerns quantum algorithms, such as Grover's algorithm, that have only demonstrated a polynomial improvement over the best known classical algorithm with respect to metrics, such as query complexity, that bear an uncertain relationship to the real physical costs of computation. 

We argue that, in order to assess the impact of such algorithms, we need a more explicitly physical model of computation. We also feel that, in order to fairly compare classical algorithms to their future quantum counterparts, we need to take into account, not just the current state of classical computing technology, but possible future developments, such as low-power reversible computing. For example, it certainly does not seem reasonable to consider extremely low cost and low power quantum memories, without assuming similar advances in classical computing technology. To this end, we have developed the Brownian computation model of Bennett, and given extensive analysis of the costs of classical and quantum algorithms for collision and preimage search.

In the case of collision search, our analysis suggests that despite their lower query complexity, quantum collision-finding algorithms do not offer a substantial, physically plausible advantage over their classical counterparts.

The case of preimage search is more delicate. In our analysis, we have developed a novel variant of Brownian computation, namely unpowered Brownian computation. It is interesting to note that, using this model of computation, we can perform a randomized classical search with the same asymptotic thermodynamic costs as Grover's algorithm. This is certainly of theoretical interest. But the practical significance of this result is somewhat less clear than in the case of collision search, since there are plausible reasons for thinking Grover's algorithm may indeed turn out to be more efficient than unpowered classical search in practice, although it should be noted that there are plausible scenarios where the reverse might hold. (As a further point of contrast, in the present state of technology, powered classical search appears to be more efficient than both approaches in finding preimages.)

We analyze in detail the technological costs which may affect the true advantage of Grover's algorithm over powered and unpowered classical preimage search. Aside from the various unique challenges involved in building fault tolerant quantum computing hardware, a key metric which appears to be relevant here is the cost of memory (or perhaps more accurately, the cost of hardware in general). As the cost of memory falls, thermodynamically reversible computing becomes more attractive to relative to current (non-reversible) computing technology. Our estimates indicate that the cost of semiconductor hardware is fairly close to the point at which reversible computing would begin to offer a real advantage. If the cost of hardware continues to fall, we would expect to see reversible computing developed for very computationally expensive tasks such as proof of work, or for very low power devices.

As the cost of memory falls further, Grover's algorithm looks less attractive, because the efficiency of classical powered search improves relative to the efficiency of Grover's algorithm, and the efficiency of unpowered classical search improves relative to the efficiency of powered classical search. Nonetheless, even without the assumption of very low hardware costs, we find that the potential advantage provided by Grover's algorithm is significantly smaller than is often assumed. Even in scenarios that are simultaneously extremely optimistic with respect to quantum computing and extremely pessimistic with regard to classical computing, Grover's algorithm will only extend the reach of classical search by a factor of $2^{46}.$ 

This analysis can be used to give guidance for post-quantum cryptography, in particular, for choosing key lengths for block ciphers. This analysis suggests that doubling the key sizes is likely unnecessary to provide protection against quantum computers, and that a smaller increase, from 128 to 192 bits for example, is likely sufficient.

\vspace{11pt}

\textit{Note:} Contributions to this work by NIST, an agency of the US government, are not subject to US copyright. Any opinions, findings, and conclusions or recommendations expressed in this material are those of the author(s), and do not necessarily reflect the views of NIST.

\bibliographystyle{plain}
\bibliography{References} 
\end{document}